# Translation and Rotation of Transformation Media under Electromagnetic Pulse

Fei Gao [1], Xihang Shi [1], Xiao Lin [1,3], Hongyi Xu [1], Baile Zhang[1,2,*]

*1. Division of Physics and Applied Physics, School of Physical and Mathematical Sciences, Nanyang Technological University, Singapore 637371, Singapore.*
*2. Centre for Disruptive Photonic Technologies, Nanyang Technological University, Singapore 637371, Singapore.*
*3. State Key Laboratory of Modern Optical Instrumentation, Zhejiang University, Hangzhou 310027, China*
*Electronic mail of corresponding author: blzhang@ntu.edu.sg

## Abstract

It is well known that optical media create artificial geometry for light, and curved geometry acts as an effective optical medium. This correspondence originates from the form invariance of Maxwell's equations, which recently has spawned a booming field called 'transformation optics.' Here we investigate responses of three transformation media under electromagnetic pulses, and find that pulse radiation can induce unbalanced net force on transformation media, which will cause translation and rotation of transformation media although their final momentum can still be zero. Therefore, the transformation media do not necessarily stay the same after an electromagnetic wave passes through.

Curved propagation of light has been found in optical media which can be generalized as effective artificial geometry for electromagnetic (EM) waves[1]. Recently, the reverse process of the generalization above has spawned a booming field named as 'transformation optics'[2-3], in which optical media with specified constitutive parameters are used to mimic prescribed distorted EM space. Essence of this field is using artificial geometry created by transformation media to replace the distorted geometry obtained by coordinate transformation[2]. The symmetry between artificial geometry and empty distorted geometry, provides an innovative design strategy of optical functional devices. This design strategy is a function-oriented approach, in which an arbitrarily distorted geometry is firstly prescribed, and then constitutive parameters of a media are calculated with transformation optics. Due to flexibility of this innovative approach, extensive interests are focused on designing various novel optical functional devices such as invisibility cloak[2-6], superlens[7], compressor[4], field rotator[8,9].

Artificial geometry has well dictated EM functions of a transformation media in controlling light, however a transformation media, as a physical object, also shows mechanical properties, not only EM functions. Due to the mechanical properties, a transformation media is not completely equivalent to artificial geometry, thus is not symmetric with curved empty geometry. Some works have investigated Lorentz force exerted on a cloak under irradiation of a monochromatic plane wave[10,11]. Spatial force distribution on a cloak is in static balance. However, few works talked about the spatio-temporal distribution of Lorentz force on a transformation media.

**Results**

Here, we will investigate the dynamic distribution of Lorentz force on three kinds of transformation media (compressor shown in Fig. 1a, carpet cloak in Fig. 1b, and rotator in Fig. 1c, whose mathematical expression will be introduced later.) under irradiation of EM pulse. In the following, Lorentz force on the three kinds of transformation media are calculated with Maxwell-Chu formula[12,13], in which macroscopic effective parameters (permittivity and permeability) are considered rather than microscopic structures[14,15]. In Chu formula, Lorentz force on media can be equivalent to four kinds of material response: electric charge density, magnetic charge density, electric current density, and magnetic current density, shown as below[12]:

$$\vec{f}(r,t) = \{(-\nabla \cdot \vec{P}(r,t)) \cdot \vec{E}(r,t) + (-\mu_0 \nabla \cdot \vec{M}(r,t)) \cdot \vec{H}(r,t) + \frac{\partial \vec{P}(r,t)}{\partial t} \times \mu_0 \vec{H}(r,t) -$$
$$-\mu_0 \frac{\partial \vec{M}(r,t)}{\partial t} \times \varepsilon_0 \vec{E}(r,t)\} \qquad (1)$$

where polarization and magnetization are defined as $\vec{P}(r,t) = (\bar{\bar{\varepsilon}} - \varepsilon_0 \bar{\bar{I}})\vec{E}(r,t)$, and $\vec{M}(r,t) = (\bar{\bar{\mu}} - \mu_0 \bar{\bar{I}})\vec{H}(r,t)$ respectively. Bound electric (magnetic) charge density (the first two terms in Eq. (1)) can be attributed to material inhomogeneity of $\bar{\bar{\varepsilon}}(\bar{\bar{\mu}})$ or longitudinal component of polarization (magnetization). As $\vec{P}$ and $\vec{H}$ in media are not parallel to each other (also for $\vec{M}$ and $\vec{E}$), contributions of current terms (the last two terms in Eq. (1)) always exist. On the other hand, a Gaussian EM pulse $\vec{H}(x,t) = \hat{z} e^{-\frac{(t-\frac{x}{c})^2}{2\sigma^2}} e^{-i\omega_0 t} e^{ik_0 x}$ is employed, where σ = 0.07, and $\omega_0$ = 5 GHz indicate width and central frequency respectively (shown in Fig. 2b). Before calculation of Lorentz force, we firstly decompose time-varying incident field into a series of time-harmonic components $\vec{H}(r,t) = \int A(\omega) \cdot \vec{H}(r,\omega) d\omega$ by Fourier transformation, where $A(\omega)$ denotes weight of each time-harmonic wave and $\vec{H}(r,\omega) = e^{-i\omega_0 t} e^{ik_0 x}$. Firstly, the monochromatic field $\vec{H}(r,\omega)$ in real space is obtained from $\vec{H}(r',\omega)$ in virtual space by coordinate transformation $(r' \to r)$. After that, the induced charge and current distributions inside transformation media are calculated. In the next by doing Fourier integration, the total field profiles are calculated as well as the total charge and current distributions. Finally, the spatio-temporal Lorentz force distribution can be calculated with Eq. (1).

We start with a simple case that a plane-wave Gaussian pulse is normally incident to a homogeneous field compressor. This slab of transformation media can be obtained by compressing transformation[4] $x' = a + x\frac{b-a}{b}$ shown in Fig. 1a, where $a$ and $b$ are left boundary and right boundary respectively. Constitutive parameters of the slab are derived as $\frac{\bar{\bar{\varepsilon}}}{\varepsilon_0} = \frac{\bar{\bar{\mu}}}{\mu_0} = diag\{\frac{b-a}{b}, \frac{b}{b-a}, \frac{b}{b-a}\}$ in rectangular coordinate, where $\varepsilon_0$ and $\mu_0$ are the permittivity and permeability of vaccum. Although this compressor is actually uniaxial, normal incidence along $\hat{x}$ direction could make coordinate frame ($\vec{E}$, $\vec{H}$, $\vec{k}$) of incident wave consistent with principle frame of $\bar{\bar{\varepsilon}}$. Therefore, this slab behaves as isotropic. As both $\vec{H}$ and $\vec{E}$ of incident wave are parallel to the two interfaces of the slab, no magnetic and electric charges are induced on the two surfaces. Thus no surface force is induced. Due to homogeneity and isotropy of compressor, the first two charge terms of Eq. (1) vanish. Considering the last two terms in Eq. (1), $\vec{P}$ and $\vec{H}$ in compressor are always perpendicular to each other as well as $\vec{M}$ and $\vec{E}$, as a result, the two current terms in Eq. (1) contribute to bulk force density. Temporal Lorentz force can be integrated as $\vec{F}(t) = \iiint \vec{f}(r,t) \cdot dV$. When the center of pulse locates at the first boundary, we let it be the time moment $t = 0$ ns. Note that the x-components of constitutive parameters $\varepsilon_x = \mu_x = \frac{b-a}{b}$ are smaller than 1, but they do not interact with the pulse. Therefore, we can consider the media non-dispersive.

The interaction process between EM pulse and compressor can be separated into five stages as in Fig. 2a. The first stage show zero velocity and acceleration ($v_x = 0$, $a_x = 0$) for the compressor, because pulse does not reach the compressor in stage one (shown in Fig. 2b at time $t = -0.17$ ns), thus zero Lorentz force. At stage two, pulse touches the left boundary but not completely enter, part of the pulse inside compressor starts to induce electric and magnetic current. Total force applied on the compressor is along $+\hat{x}$ direction and reaches maximum (Fig. 2a) at $t = 0$ ns (in Fig. 2c). Therefore compressor is moving and accelerating along $+\hat{x}$ direction with ($v_x > 0$, $a_x > 0$). The oscillation of Lorentz force in stage two in Fig. 2a is not symmetric. That is because the velocity of pulse inside compressor is smaller than that in background. When pulse completely enter the compressor, as stage three starts, the force density distribution is symmetric with respect to the pulse center (shown in Fig. 2d), so the total force decreases to zero. In spite of zero total Lorentz force, according to impulse theorem, the compressor keeps a constant velocity along $+\hat{x}$ direction in stage three ($v_x > 0$, $a_x = 0$), due to the positive Lorentz force in stage two. The force distribution of stage four (when pulse reaches the right boundary) is opposite to that of stage two, and the total force is along $-\hat{x}$. The compressor begins to decelerate. Stage five begins when the pulse completely leaves compressor (shown in Fig. 2e at time $t = 0.67$ ns). Lorentz force at this stage is zero. Due to above processes, the compressor gets net translation but its velocity gets back to zero. Note that we here utilize quasi-static assumption in calculating Lorentz force, because velocity of macroscopic transformation media is negligible compared with light speed. Generally, the pulse firstly transfers momentum to the compressor and then takes it back. As a result, the whole process is dynamically balanced.

In the following, we include anisotropy to study Lorentz force on a transformation media, so we choose the lifting transformation (shown in Fig. 1b) used in the design of carpet cloak which is placed on a smooth perfect electric conductor (PEC) (indicated with tilted dashes in Fig. 1b). Following the transformation formula in Ref. 6, we can get constitutive parameters in the triangular region enclosed by CBD as $\frac{\bar{\bar{\varepsilon}}}{\varepsilon_0} = \begin{bmatrix} 1/\kappa & -\tau/\kappa \\ -\tau/\kappa & \kappa + \tau^2/\kappa \end{bmatrix}, \frac{\bar{\bar{\mu}}}{\mu_0} = 1/\kappa$, which is mirror symmetric with that in the triangular region enclosed by ACD with respect to the vertical plane, where $\varepsilon_0$ and $\mu_0$ are relative permittivity and permeability of background material, and $\tau = tan\beta$, $\kappa = \frac{\tan(\alpha+\beta)-tan\beta}{\tan(\alpha+\beta)}$, where $\alpha = 20.01°$ and $\beta = 21.8°$ are two angles indicated in Fig. 3b. Although this carpet cloak is homogeneous, off-diagonal term $-\frac{\tau}{\kappa}$ in $\bar{\bar{\varepsilon}}$ could induce longitudinal polarization field component, thus bulk bound electric charge density. For a single frequency, the charge density is

$$\rho_e(\omega) = real(i\frac{k^2\tau}{\omega\kappa}e^{ikx-i\omega t}) \qquad (2)$$

in the triangular region enclosed by CBD, where $k$ is the wavevector in the background material. Temporal bulk charge density under the same pulse in compressor case can be obtained with $\rho_e(t) = \int A(\omega) \cdot \rho_e(\omega) d\omega$. As $\bar{\bar{\mu}}$ is homogeneous and isotropic, so magnetic charges $-\mu_0 \nabla \cdot \vec{M}(r,t)$ does not contribute. Contributions of the two current terms are similar with the analysis in compressor. Different from the situation in compressor, electric field here is not parallel with the surfaces of this carpet cloak, so surface bound charge density exist on the surface ACB and ADB. Thus the surface Lorentz force exists. On surface ACB, the electric charge term $-\nabla \cdot \vec{P}$ is reduced to $\hat{n} \cdot \varepsilon_0 (\vec{E}_{background} - \vec{E}_{cloak})$[10], where $\hat{n}$ denotes the normal vector of surface ACB from cloak to background. The electric field on ACB is obtained by taking the average of $\vec{E}$ on both sides of ACB. Note that the causality issue can be addressed by immersing the whole cloak into a higher-index background media. But in that scenario Lorentz force can distribute in the whole space. Here we only consider the Lorentz force on the cloak.

Calculated surface Lorentz force on ACB is shown in Fig. 3a. Besides surface charges, surface electric current[11] on PEC also contribute to the surface Lorentz force on ADB. With respect to the surface Lorentz force on ACB, that on ADB is a little smaller and with opposite direction (in Fig. 3a). By similar calculation with above, the force on interface CD is along $\hat{y}$ direction. However, the force along $\hat{y}$ direction cannot induce any movement, because of gravity force of cloak. Fig. 3a shows the time evolution of total Lorentz force along $\hat{x}$ direction, provided that pulse locates in the middle of cloak at $t = 0$ ns (shown in Fig. 3c). The dynamic process can be divided into two parts when the pulse propagates through the cloak. From $t = -1.66$ ns to $t = 0$ ns, total force (red solid in Fig. 3a) exerted on cloak is positive, so the cloak accelerates along positive $\hat{x}$ direction. Pulse pattern in the first stage is shown at $t = -0.33$ ns in Fig. 3b. Deceleration occurs during 0 ns ~ 1.66 ns, and the cloak continuingly moves along $\hat{x}$ direction till zero velocity at $t = 1.66$ ns. Pulse pattern in stage two is shown at $t = 0.67$ ns in Fig. 3d. The total force is dynamically balanced, and net translation of the cloak along $\hat{x}$ occurs.

Finally, we will further include inhomogeneity to study spatio-temporal Lorentz force density on a transformation media. A field rotator[8] is employed here, and the rotation transformation is shown in Fig. 1c. Its permitivity and permeability are expressed in cylindrical coordinate as $\frac{\bar{\bar{\varepsilon}}}{\varepsilon_0} = \frac{\bar{\bar{\mu}}}{\mu_0} = \begin{bmatrix} 1 & -t & 0 \\ -t & t^2+1 & 0 \\ 0 & 0 & 1 \end{bmatrix}$, where $t = \frac{r\theta_0}{R_2-R_1}$, $R_2$ and $R_1$ are outer and inner radius respectively. $r$ is radius of arbitrary point and $\theta_0$ is arbitrary rotating angle. Here, we take $\theta_0 = \frac{\pi}{2}$, and $R_2 = R_1 = 5\lambda$, where $\lambda$ is the wavelength in vaccum. Since $\vec{E}$ field is in $x$-$y$ plane, both the off-diagonal term of $\bar{\bar{\varepsilon}}$ and material inhomogeneity will contribute to electric charge density. For a single frequency, bulk electric charge density is as follow:

$$\rho_e(\omega) = \frac{k}{\omega\rho} real\{e^{ik\rho cos\theta - i\omega t}[3tcos\theta - t^2 sin\theta + ik\rho(t \cdot cos^2\theta - t \cdot sin^2\theta - t^2 sin\theta cos\theta)]\} \tag{3}$$

where $k$ is the wavevector in vaccum, and ($\rho$, $\theta$) is cylindrical coordinate of arbitrary point. We specify that when $t = 0$ ns, the pulse center is located at the middle of the rotator. In this scenario, Lorentz force is in $x$-$y$ plane, but not along a specific direction, so for simplicity we calculate torque $\vec{m}_v = \vec{r} \times \vec{f}$ with respect to rotator center. Fig.

4a shows the evolution of total torque on the rotator. Bulk and surface torque are also separately shown in Fig. 4a. It is obvious that the bulk region contributes most to the total torque, and surface torque on the inner surface can be neglected. Here the causality issue can be argued with the same reason as that of the carpet cloak discussed above.

The whole process, when pulse go through the rotator, can be divided into four stages. In stage one (-1.66 ns < $t$ < -0.8 ns), angular momentum transferred from pulse drives rotator to acceleratingly rotate clockwisely (angular velocity $\omega$ < 0, angular acceleration $a$ < 0). Then, in stage two (-0.8 ns < $t$ < 0 ns), the rotator deceleratingly rotates clockwisely ($\omega$ < 0, $a$ > 0). Corresponding field pattern at $t$ = 0.67 ns is shown in Fig. 4b. Further deceleration leads rotator to a critical point $t$ = 0 ns (field pattern shown in Fig. 4c), and the rotator is static at that time. After the critical point, the rotator enters into stage three (0 ns < $t$ < 0.8 ns), and starts to acceleratingly rotate anti-clockwisely ($\omega$ > 0, $a$ > 0). Till $t$ = 0.8 ns, angular velocity $\omega$ reaches maximum, while acceleration $a$ vanishes. After that, stage four (0.8 ns < $t$ < 1.66 ns) begins, and corresponding field pattern at $t$ = 1.0 ns is shown in Fig. 4d. In this stage, the rotator deceleratingly rotate anti-clockwisely ($\omega$ > 0, $a$ < 0). The integration of torque over time is zero, which means the total angular momentum transferred to rotator is zero, i.e. the whole process is dynamically balanced.

**Conclusion**

In conclusion, our results show that translation of compressor, carpet cloak and rotation of rotator can be observed under incidence of EM pulse respectively. In the whole process of pulse propagation, total force or torque on transformation media is eventually balanced, but can be nonzero for a short time period. The motion of transformation media verifies the asymmetry between transformation media and curved geometry. For demonstration of this asymmetry, transformation media can be realized with metamaterials consisting of subwavelength microstructures. Furthermore, the motion process of transformation media can be engineered with concrete microstructure design[14,15], because different microstructures, as realizations of the same macro effective parameters, can generate different force distributions.

**Acknowledgement**

We thank Prof. Brandon A. Kemp from Arkansas State University for beneficial discussion. This work was sponsored by the NTU Start-Up Grants, Singapore Ministry of Education under Grant No. MOE2015-T2-1-070 and MOE2011-T3-1-005.

**Authors Contributions**

F.G., performed the calculation. X.S., X.L., and H.X. involved in preparing figures. B.Z. supervised the project.

**Additional Information**

Competing financial interests: the authors declare no competing financial interests.

**References**:

1. Leonhardt, U. & Philbin, T.G. Transformation Optics and the Geometry of Light. *Progress in Optics* **53**, 69-152 (2009).
2. Pendry, J. B. *et al.* Controlling electromagnetic fields. *Science* **312**, 1780-1782 (2006).
3. Leonhardt, U. Optical conformal mapping. *Science* **312**, 1777-1780 (2006).
4. Pendry, J. B. Taking the wraps off cloaking. *Physics* **2**, 95 (2009).
5. Schurig, D. *et al.* Metamaterial electromagnetic cloak at microwave frequencies. *Science* **314**, 977-980 (2006).
6. Zhang, B.L. *et al.* Macroscopic Invisibility Cloak for Visible Light. *Phys. Rev. Lett.* **106**, 3 (2011).
7. Pendry, J.B. Negative refraction makes a perfect lens. *Phys. Rev. Lett.* **85**, 3966-3969 (2000).
8. Chen, H.Y. & Chan, C. T. Transformation media that rotate electromagnetic fields. *Appl. Phys. Lett.* **90**, 24 (2007).
9. Chen, H. Y. *et al*. Design and Experimental Realization of a Broadband Transformation Media Field Rotator at Microwave Frequencies. *Phys. Rev. Lett.* **102**, 18 (2009).
10. Chen, H. S. *et al.* Lorentz force and radiation pressure on a spherical cloak. *Phys. Rev. A*. **80**, 1 (2009).
11. Chen, H. S. *et al.* Optical force on a cylindrical cloak under arbitrary wave illumination. *Opt. Lett.* **35**, 667-669 (2010).
12. Kemp, B.A., Resolution of the Abraham-Minkowski debate: Implications for the electromagnetic wave theory of light in matter. *J. Appl. Phys.* **109**, 11 (2011).
13. Kong, J.A. *Electromagnetic Wave Theory*, Cambridge, MA: EMW Publishing (2008).
14. Wang, S., Ng, J., Xiao, M., Chan, C. T. Electromagnetic stress at the boundary: Photon pressure or tension? *Sci. Adv.* **2**, e1501485 (2016).
15. Kemp, B. A. Nanophotonics: Momentum in metamaterials. *Nature Photon.* **10**, 291-293 (2016).

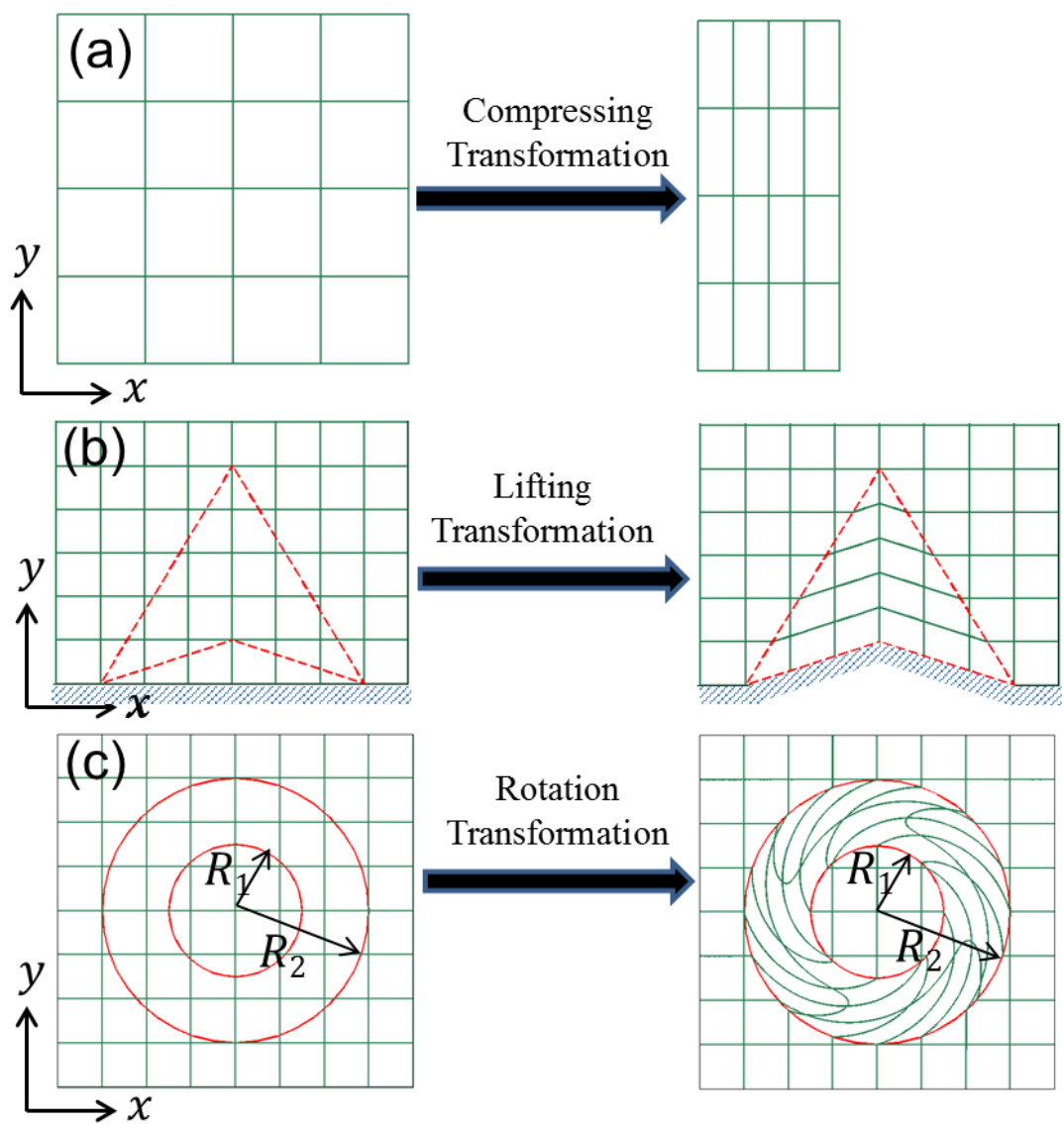

**Figure 1.** (**a**) shows compressing transformation from left to right, (**b**) shows lifting transformation of carpet cloak, (**c**) shows rotation transformation from left to right.

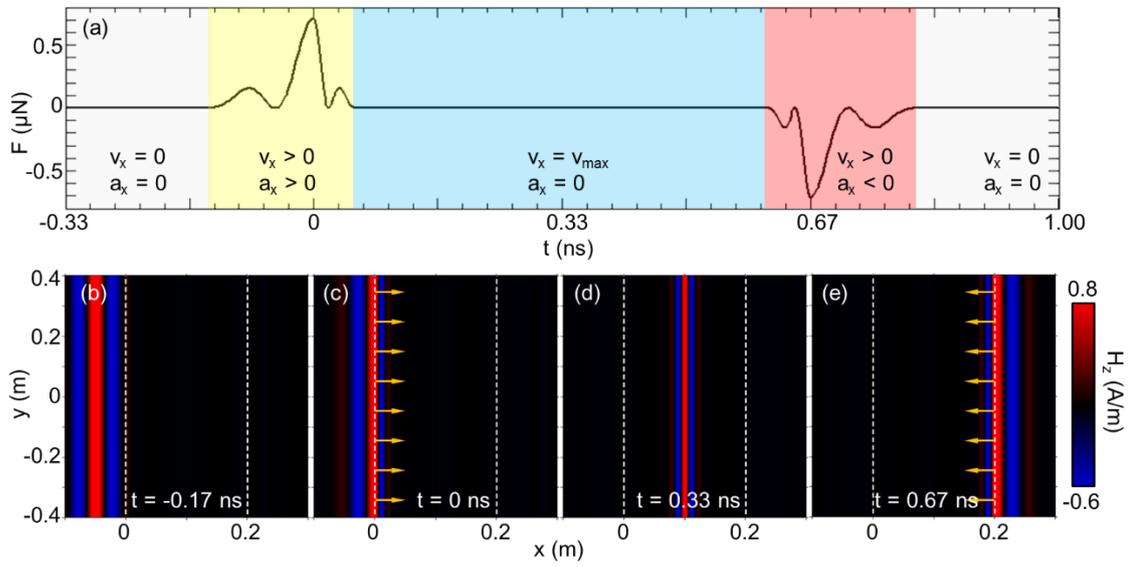

**Figure 2. (a)** shows total force $F(t)$ applied on compressor versus time, $a_x$ and $v_x$ represent acceleration and speed along $\hat{x}$ direction respectively, **(b)**-**(e)** show $H_z$ component of pulse with color at different times and arrows represent total Lorentz force $F$ on the compressor. The two vertical dashed lines denote the boundaries of the compressor.

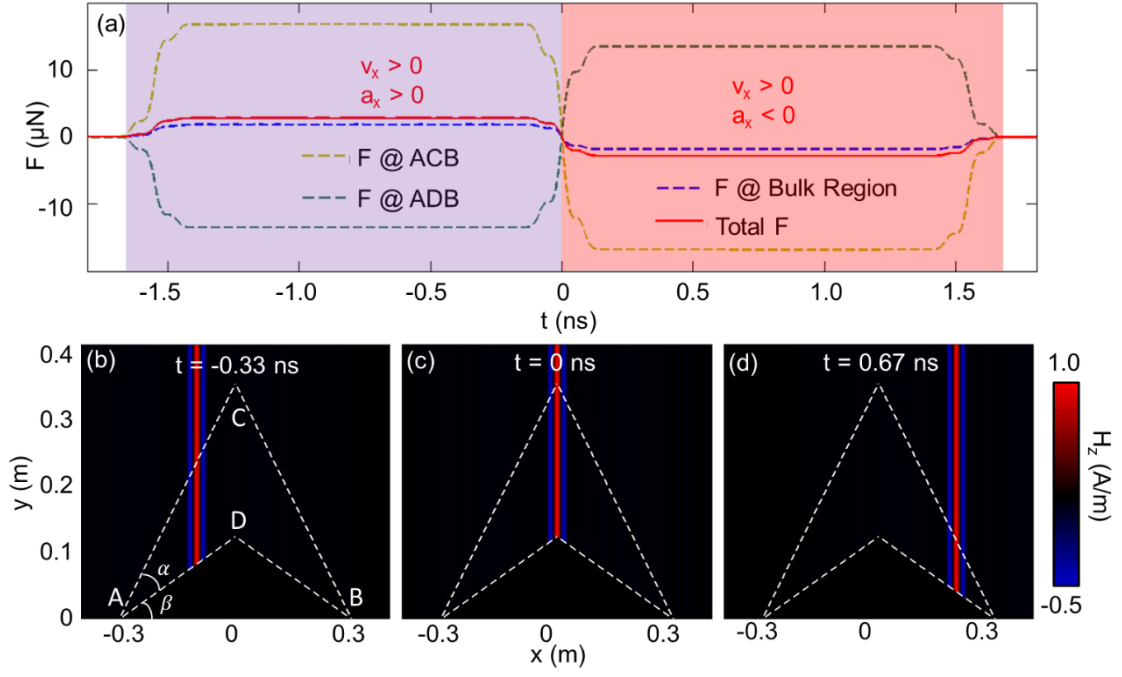

**Figure 3. (a)** shows total force $F(t)$ applied on carpet cloak versus time, $a_x$ and $v_x$ represent acceleration and speed along $\hat{x}$ direction respectively, **(b)**-**(d)** show $H_z$ component of pulse with color at different times. Region enclosed by dashed lines is the carpet cloak.

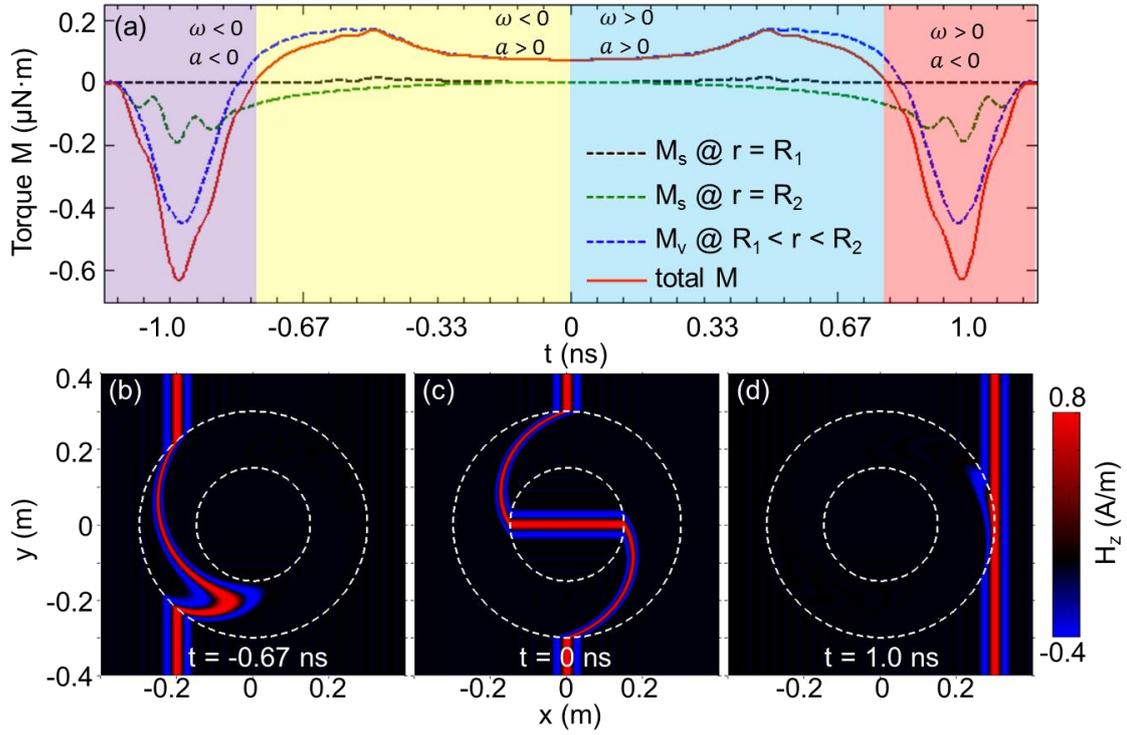

**Figure 4.** (a) shows torque $M$ of different region versus time, dark dashed, green dashed, blue dashed and red solid curves represent inner surface, outer surface, bulk and total torque respectively. $\omega$ and $a$ represent angular velocity and angular acceleration respectively, (b)-(d) show $H_z$ component of pulse with color at different times. The two dashed circles represent inner and outer boundaries respectively.